\documentclass[12pt]{article}
\usepackage{amsmath,amssymb,latexsym}

\begin{document}

\begin{titlepage}
\noindent

\begin{flushright}
	YITP-01-31 \\
	hep-th/0105002
\end{flushright}

\vspace{2.5cm}

\begin{center}
{\Large\bf A New Approach to}\\
\vspace{0.5cm}
{\Large\bf Scalar Field Theory on Noncommutative Space}\\
\bigskip
\vspace{1.5cm}
{\large Yoshinobu Habara}~\footnote{e-mail: habara@yukawa.kyoto-u.ac.jp}\\
\vspace{0.5cm}
{\it  Yukawa Institute for Theoretical Physics,}\\
{\it  Kyoto University, Kyoto 606-8502, Japan}
\end{center}

\vspace{1.5cm}

\begin{abstract}
\noindent

A new approach to constructing the noncommutative scalar field theory is presented. Not only between $\hat{x}_i$ and $\hat{p}_j$, we impose commutation relations between $\hat{x}_i$s as well as $\hat{p}_j$s, and give a new representation of $\hat{x}_i,\hat{p}_j$s. We carry out both first- and second-quantization explicitly. The second-quantization is performed in both the operator formalism and the functional integral one.

\end{abstract}

\vspace{0.5cm}

\end{titlepage}

\noindent
{\bf 1. Introduction}

\vskip 0.5cm

Recently, the extended geometry known as noncommutative geometry is widely discussed in the contexts of field theories, string theories, Matrix models (Ref.~\cite{schwarz}~\cite{susskind}~\cite{nekrasov}~\cite{witten} etc.). Mathematically (Ref.~\cite{connes}~\cite{landi} etc.), the ``point" $x$ in noncommutative space is regarded as the operator $\hat{x}$ acting on a Hibert space and the space coordinates as the eigenvalues of $\hat{x}$. Until now, in field theories, $*$-product has been used which puts the noncommutativity only to the product between the usual numbers instead of the operators $\hat{x}$. The field theories following this tactics contains the very interesting problems such as UV/IR mixing(Ref.~\cite{seiberg} etc.), while they have divergence of the same degree as the usual field theories. Therefore, using $*$-product as noncommutativity, I believe that one may not be able to obtain consistent quantum gravity model. 

In this paper, I would like to propose a new approach to introduce the noncommutativity into the scalar field theory. Our approach is based on the argument that first-quantization is equal to the introduction of noncommutativity into the phase space and that the first-quantized theory to the classical field theory. The quantum mechanics discussed until now have noncommutativity only between the space-time coordinate $x_i$ and the momentum $p_j$ like $[\hat{x_i},\hat{p_j}]=i\delta_{ij}$, and the representation of this algebra is defined by $\hat{x}_i=x_i,\; \hat{p}_j=-i\frac{\partial}{\partial x_j}$. But I would like to also impose noncommutativity between $x_i$s as well as $p_j$s, and make a new representation of $\hat{x}_i$s and $\hat{p}_j$s. This new representation leads us to a new Klein-Gordon model with discrete spectrum which can be exactly solved using the harmonic oscillator system in ``Euclidean" case. Also, we will carry out second-quantization in both the operator formalism and the functional integral formalism. In the later sections, we will find a slightly strange creation-annihilation algebra of fields and functional integral. This functional integral has special characteristics such as the summation $\sum_{N,l}$ instead of the momentum-integration $\int d^nk$ and the propagator with imarinary part looking like the regulator.

The organization of this paper is as follows. In section 2, we present a noncommutative algebra and it's representation. In section 3, we solve Klein-Gordon equation on noncommutative space as first-quantized theory. Section 4 and 5 are devoted to second-quantization by the operator formalism and the functional integral formalism respectively. In section 6, discussions are presented.

\vskip 1.5cm

\noindent
{\bf 2. Representation of Euclidean noncommutative spaces}

\vskip 0.5cm

First of all, we will give an algebra of the ``Euclidean" noncommutative $\textbf{R}^n$ and its conjugate momenta as the first quantized theory.

The commutation relations between the space (or space-time) coordinates $\hat{x}_i$s are at the order of $\theta$ where $\theta$ denotes the noncommutativity of the space-time, and those between $\hat{x}_i$ and it's conjugate momentum $\hat{p}_i$ are the canonical commutation relations.

\begin{eqnarray}
	& [\hat{x}_i ,\hat{x}_j ] & \! \! \! \! =i\theta_{ij} \nonumber \\
	& [\hat{x}_i ,\hat{p}_j ] & \! \! \! \! =i\hbar \delta_{ij} \qquad 
	(i,j=1\sim n) \\
	& [\hat{p}_i ,\hat{p}_j ] & \! \! \! \! =\> ?? \nonumber
\end{eqnarray}

We call this algebra $\mathcal{A}$. Though the commutation relations between $\hat{p}_i$s will be exactly determined later, we shall roughly evaluate them here.

Because of these commutation relations, the deviations $\Delta x$ and $\Delta p$ have the following uncertainty relations.

\begin{eqnarray*}
	& [\hat{x}_i ,\hat{x}_j ] & \! \! \! \! =i\theta_{ij} \quad \Rightarrow
	 \quad (\Delta x)^2 \geq \theta \\
	& [\hat{x}_i ,\hat{p}_j ] & \! \! \! \! =i\hbar \delta_{ij} \quad 
	\Rightarrow \quad \Delta x \Delta p \geq \hbar \text{,} \quad 
	(\Delta x)^2(\Delta p)^2 \geq {\hbar}^2
\end{eqnarray*}

Thus,

\begin{eqnarray*}
	[\hat{p}_i ,\hat{p}_j ] \propto {\hbar}^2 {\theta}^{-1}_{ij}
\end{eqnarray*}

In the following discussion, we will take $\hbar=1$.

Let us describe $\hat{x}_i$s and $\hat{p}_i$s directly using the operators in Ref.~\cite{schwarz}. Define the operators $\hat{U}_i$ as follows :

\begin{eqnarray}
	& & \hat{U}_i \equiv e^{i \hat{x}_i} \qquad (i,j=1\sim n)\nonumber \\
	& & \hat{U}_i \hat{U}_j =e^{-i\theta_{ij}} \hat{U}_j \hat{U}_i
\end{eqnarray}

$\hat{U}_i$s are the operators acting on Hilbert space $\mathcal{H} \ni f(t)$, where $t=(t_1,\dots ,t_n) \in \textbf{R}^n$. The representation of $\hat{U}_i$s satisfing the commutation relations (2) can be constructed in the following way. We begin by supposing the following form of $\hat{U}_i$ acting on $\mathcal{H}$ : 

\begin{eqnarray}
	& (\hat{U}_i f) \! \! \! \! & (t)=\chi_i(t)f(t+a_i) \\
	& & \chi_i : \textbf{R}^n \rightarrow \textbf{C} \quad 
	\text{(character homomorphism)} \nonumber
\end{eqnarray}

Using the commutation relations (2), we obtain

\begin{eqnarray}
	& \hat{U}_i \hat{U}_j f(t)=\chi_j(a_i) \chi_i(t) \chi_j(t) 
	f(t+a_i+a_j) \nonumber \\
	& \hat{U}_j \hat{U}_i f(t)=\chi_i(a_j) \chi_i(t) \chi_j(t) 
	f(t+a_i+a_j) \nonumber \\
	& \text{thus,} \quad \displaystyle e^{-i\theta_{ij}}=
	\frac{\chi_j(a_i)}{\chi_i(a_j)} \text{.}
\end{eqnarray}

Here we can take

\begin{eqnarray*}
	a_i=-\theta_{ij}z_j \text{,} \qquad z_j=(0,\dots 
	,\! \! \! \! & 1 & \! \! \! \! ,\dots ,0) \text{,} \\
	& \widehat{\text{\tiny $j$}} & \\
	\chi_i(t)=e^{\frac{i}{2}t_i} \text{.} & &
\end{eqnarray*}

So the representation of ${U}_i$s are given by

\begin{equation}
	\hat{U}_i f(t)=e^{\frac{i}{2}t_i} f(t-\theta_{ij}z_j) \text{.}
\end{equation}

Next, in order to show the representation of $\hat{x}_i$, we take

\begin{eqnarray*}
	& & \hat{U}_i =e^{ik\hat{x}_i} \\
	& & z_i=(0,\dots ,k,\dots ,0) \\
	& & \chi_i(t)=e^{\frac{i}{2}kt_i} \text{.}
\end{eqnarray*}

Substituting them into (5), we expand in $k$, and compare the both side at the first order of $k$ to obtain

\begin{align*}
	e^{ik\hat{x}_1} f(t_1,t_2,\dots ,t_n) &=e^{\frac{i}{2}kt_1} 
	f(t_1,t_2-k\theta_{12},\dots ,t_n-k\theta_{1n}) \\
	(1+ik\hat{x}_1+\cdot \cdot \cdot) f(t_1 &,t_2,\dots ,t_n) \\
	&=(1+\frac{i}{2}kt_1+\cdot \cdot \cdot)\{ 1-k(\theta_{12}\partial_2
	+\dots +\theta_{1n}\partial_n)+\cdot \cdot \cdot \} 
	f(t_1,t_2,\dots ,t_n) \\
	\Rightarrow \hat{x}_1= \frac{1}{2}t_1 &+i\theta_{1j}\partial_j 
	\text{,} \qquad \text{where } \; \partial_i 
	=\frac{\partial}{\partial t_i}
\end{align*}

\noindent and thus,

\begin{eqnarray}
	\hat{x}_i=\frac{1}{2}t_i+i\theta_{ij}\partial_j \text{.}
\end{eqnarray}

Finaly, we determine the representation and the algebra of $\hat{p}_i$s. In the similar manner to the former construction, we use the operators $\hat{U}_i$. From the canonical commutation relations $[\hat{x}_i, \hat{p}_j]=i\delta_{ij}$, we have

\begin{eqnarray*}
	[e^{i\hat{x}_i}, \hat{p}_j]=-e^{i\hat{x}_i}\delta_{ij} \text{,}
\end{eqnarray*}

\noindent thus,

\begin{eqnarray*}
	\hat{p}_i +\delta_{ij}=\hat{U}^{-1}_j \hat{p}_i \hat{U}_j \text{.}
\end{eqnarray*}

If we set $\; \hat{p}_if(t)=A_{ij}t_jf(t)+B_{ij}\partial_jf(t)$,

\begin{eqnarray*}
	& & (\hat{p}_i +\delta_{ij})f(t)=(A_{ik}t_k+B_{ik}\partial_k
	+\delta_{ij})f(t) \\
	& & \hat{U}^{-1}_j \hat{p}_i \hat{U}_j f(t)=\{ A_{ik}(t_k+\theta_{jl}
	z_l)+\frac{i}{2}B_{ij}+B_{ik}\partial_k\} f(t) \text{.}
\end{eqnarray*}

Thus,

\begin{eqnarray*}
	\delta_{ij}=A_{ik}\theta_{jk}+\frac{i}{2}B_{ij}
	=-A_{ik}\theta_{kj}+\frac{i}{2}B_{ij} \text{.}
\end{eqnarray*}

Therefore we can choose

\begin{eqnarray*}
	& & A_{ij}=-\frac{1}{2}\theta^{-1}_{ij} \text{,} \quad 
	B_{ij}=-i\delta_{ij} \text{.}
\end{eqnarray*}

So that

\begin{eqnarray}
	& & \hat{p}_i=-\frac{1}{2}\theta^{-1}_{ij}t_j
	-i\partial_i \\
	& & \quad \Rightarrow [\hat{p}_i ,\hat{p}_j ]=-i\theta^{-1}_{ij} 
	\text{.} \nonumber
\end{eqnarray}

There may be other choice of $A_{ij}$ and $B_{ij}$, but the other choices will lead to the different representations and algebras. However, in order to satisfy the relation $(\Delta p)^2 \geq \theta^{-1}$, the variations are restricted.

If we would like to take the commutative space limit $\theta \! \to \! 0$, we must keep in mind the following relation.

\begin{eqnarray*}
	\hbar^2 \ll \theta \ll \hbar
\end{eqnarray*}

Under this inequality, we have the canonical commutation relations and representations, i.e. quantum mechanics on commutative space.

In this representation, there are relations between $\hat{x}_i$s and $\hat{p}_i$s like 

\begin{eqnarray*}
	\hat{p}_i =-\theta^{-1}_{ij} \hat{x}_j \text{,}
\end{eqnarray*}

\noindent but we have $2n$ canonical ``independent" operators $t_i$s and $\partial_i$s. This may mean that the classical spase $(x_1, \dots ,x_n)$ is different from the quantum space $(t_1, \dots ,t_n)$ by $\theta$ in our current view that first-quantization is equal to the introduction of noncommutativity into the phase space.

\vskip 1.5cm

\noindent
{\bf 3. First quantization : deformed Klein-Gordon theory}

\vskip 0.5cm

The purpose of this section is to solve the Klein-Gordon model on the noncommutative space. For simplicity, we will take the dimension of the space to be $2$, and the ``metric" is taken to be Euclidean.

The Klein-Gordon model, the relativistic first-quantized theory, is constructed from the Einstein's energy-momentum relation $p_ip_i+m^2=0$ by replacing the phase-space coordinates with the operators : $x_i\text{,} \; p_i \rightarrow \hat{x}_i\text{,} \; \hat{p}_i$. On noncommutative space, we use the same method.

We start with the following Klein-Gordon equation : 

\begin{eqnarray}
	& & (\hat{p}_i \hat{p}_i +m^2)\phi (t)=0 \text{,} \\
	& & [(-\frac{1}{2}\theta^{-1}_{ij}t_j-i\partial_i)
	(-\frac{1}{2}\theta^{-1}_{ik}t_k-i\partial_k)+m^2]\phi (t)=0 \text{.}
\end{eqnarray}

Our space is $2$-dimensional, so $i,j,k=1\sim 2$ and we can take the $\theta_{ij}$'s to be

\begin{equation}
	\theta_{ij}=\theta \left(
	\begin{array}{cc}
	0 & 1 \\
	-1 & 0
	\end{array}
	\right)
	, \quad 
	\theta^{-1}_{ij}=\theta^{-1} \left(
	\begin{array}{cc}
	0 & -1 \\
	1 & 0
	\end{array}
	\right)
\end{equation}

Let us describe the operators and equations explicitly in the following : 

\begin{eqnarray}
	& & \left\{
	\begin{array}{l}
	\hat{x}_1=\frac{1}{2}t_1+i\theta \partial_2 \\
	\hat{x}_2=\frac{1}{2}t_2-i\theta \partial_1 
	\end{array} \right.
	, \quad 
	\left\{
	\begin{array}{l}
	\hat{p}_1=\frac{1}{2\theta}t_2-i\partial_1 \\
	\hat{p}_2=-\frac{1}{2\theta}t_1-i\partial_2 
	\end{array} \right. \\
	& & \nonumber \\
	& & (\hat{p}_1^2+\hat{p}_2^2+m^2)\phi(t_1,t_2)=0 \\
	& & \Big[ -(\partial_1^2+\partial_2^2)+\frac{1}{4\theta^2}(t_1^2+t_2^2)
	+\frac{i}{\theta}(t_1\partial_2-t_2\partial_1)+m^2\Big] \phi(t_1,t_2)=0 	\text{.}
\end{eqnarray}

The field $\phi(t_1,t_2)$ belongs to the Hilbert space $\mathcal{H}\equiv L^2(\textbf{R}^2)$ (space of square-integrable functions).

The operator part of the Klein-Gordon equation (13) is of the form : 

\begin{equation*}
	\ll \text{$2$-dimensional harmonic oscillator}\gg+\ll
	\text{angular momentum}\gg +m^2 \text{.}
\end{equation*}

The origin of the name ``angular momentum" will be appeared later. Equation (13) is of the same form as that of electron in the magnetic field for the discussion of Landau level. Before (13), we can consider that scalar field exists in the presence of gauge field $A_i(t)=\frac{1}{2} \theta^{-1}_{ij}t_j$ in view of equation (9).

In this system, tha harmonic oscillator is contained. So we introduce the creation-annihilation operators as usual,

\begin{eqnarray}
	& & \left\{
	\begin{array}{l}
	\hat{a}_1=\sqrt{\theta}(\frac{1}{2\theta}t_1+\partial_1) \\
	\hat{a}_1^{\dagger}=\sqrt{\theta}(\frac{1}{2\theta}t_1-\partial_1) 
	\end{array} \right.
	, \quad 
	\left\{
	\begin{array}{l}
	\hat{a}_2=\sqrt{\theta}(\frac{1}{2\theta}t_2+\partial_2) \\
	\hat{a}_2^{\dagger}=\sqrt{\theta}(\frac{1}{2\theta}t_2-\partial_2) 
	\end{array} \right. \\
	& & [\hat{a}_i, \hat{a}_j^{\dagger}]=\delta_{ij}, \quad 
	[\hat{a}_i, \hat{a}_j]=[\hat{a}_i^{\dagger}, \hat{a}_j^{\dagger}]=0 
	\text{.}
\end{eqnarray}

Then, $\ll \! 2$-dimensional harmonic oscillator$\gg$ part and $\ll$angular momentum$\gg$ part become the following form respectively.

\begin{eqnarray}
	\ll \text{$2$-dimensional harmonic} & & \! \! \! \! \! \! \! 
	\text{oscillator}\gg \nonumber \\
	\hat{H}_h & & \! \! \! \! \! \! \! =-(\partial_1^2+\partial_2^2)
	+\frac{1}{4\theta^2}(t_1^2+t_2^2) \qquad \qquad \qquad \nonumber \\
	& & \! \! \! \! \! \! \! =\frac{1}{\theta}(\hat{a}_1^{\dagger}\hat{a}_1
	+\hat{a}_2^{\dagger}\hat{a}_2+1) \nonumber \\
	& & \! \! \! \! \! \! \! \equiv \frac{1}{\theta}(\hat{N}+1) \\
	\ll \text{angular momentum} & & \! \! \! \! \! \! \! \gg \nonumber \\
	\frac{1}{\theta} \hat{L} & & \! \! \! \! \! \! \! \equiv 
	\frac{i}{\theta}(t_1\partial_2-t_2\partial_1) \nonumber \\
	& & \! \! \! \! \! \! \! =\frac{i}{\theta}(\hat{a}_1^{\dagger}\hat{a}_2
	-\hat{a}_1\hat{a}_2^{\dagger})
\end{eqnarray}

Here, by simple calculation, we see that

\begin{equation}
	[\hat{N}, \hat{L}]=0 \text{.}
\end{equation}

Thus, $\hat{N}$ and $\hat{L}$ are simultaneously diagonalizable so that we can write the eigenstates as $|N,l\rangle$ with the eigenvalues $(N,l)$ of $(\hat{N},\hat{L})$.

\begin{eqnarray}
	& & \frac{1}{\theta}(\hat{N}+1)|N,l\rangle =\frac{1}{\theta}(N+1)
	|N,l\rangle \\
	& & \frac{1}{\theta}\hat{L}|N,l\rangle =\frac{1}{\theta}l|N,l\rangle
\end{eqnarray}

From the equation of motion (13), we get the mass-shell condition :

\begin{equation}
	\frac{1}{\theta}(N+1)+\frac{1}{\theta}l+m^2=0 \text{.}
\end{equation}

Next, we will write down the eigenstates $|N,l\rangle$ explicitly. Since the eigenvalue $N$ is the one of $2$-dimensional harmonic oscillator, $|N,l\rangle$ can be decomposed into the two $1$-dimensional harmonic oscillators as follows : 

\begin{eqnarray*}
	|N,l\rangle & \! \! \! = & \! \! \! \sum_{n=0}^N c_n^{(N)}(l) 
	|n,N-n\rangle \\
	& \! \! \! = & \! \! \! \sum_{n=0}^N c_n^{(N)}(l) |n\rangle \otimes 
	|N-n\rangle \text{.}
\end{eqnarray*}

By substituting this decomposition into (20), we can determine the coefficients $c_n^{(N)}(l)$.

\begin{eqnarray*}
	\hat{L}|N,l\rangle & \! \! \! = \! \! \! & i(\hat{a}_1^{\dagger}
	\hat{a}_2-\hat{a}_1 \hat{a}_2^{\dagger}) \sum_{n=0}^N c_n^{(N)}(l) 
	|n,N-n\rangle \\
	& \! \! \! = & \! \! \! i\sum_{n=1}^N c_{n-1}^{(N)}(l) \sqrt{n(N-n+1)}
	|n,N-n\rangle \\
	& \! \! \! & \qquad \qquad -i\sum_{n=0}^{N-1} c_{n+1}^{(N)}(l) 
	\sqrt{(n+1)(N-n)}|n,N-n\rangle \\
	& \! \! \! = & \! \! \! l\sum_{n=0}^N c_n^{(N)}(l) |n,N-n\rangle \\
	\text{\small $(n=0)$} & & -ic_1^{(N)}(l) \sqrt{N}=lc_0^{(N)}(l) \\
	\text{\small $(1 \leq n \leq N-1)$} & & ic_{n-1}^{(N)}(l) 
	\sqrt{n(N-n+1)}-ic_{n+1}^{(N)}(l) \sqrt{(n+1)(N-n)} =lc_n^{(N)}(l) \\
	\text{\small $(n=N)$} & & ic_{n-1}^{(N)}(l) \sqrt{N}=lc_n^{(N)}(l)
\end{eqnarray*}

All $c_n^{(N)}(l)$s, $(1\leq n\leq N)$ are described with $c_0^{(N)}(l)$ where $c_0^{(N)}(l)$ is determined by normalizing $|N,l\rangle$.

Now, we shall discuss the eigenvalue $l$ of $\hat{L}$ in detail. We can show immediately the following commutation relations.

\begin{eqnarray*}
	& & \left\{
	\begin{array}{ll}
	\! \! \! \! \! \! & [\hat{L}, \hat{a}_1^{\dagger}\hat{a}_1
	-\hat{a}_2^{\dagger}\hat{a}_2]=-2i(\hat{a}_1^{\dagger}\hat{a}_2
	+\hat{a}_1\hat{a}_2^{\dagger}) \\
	\! \! \! \! \! \! & [\hat{L}, \hat{a}_1^{\dagger}\hat{a}_2
	+\hat{a}_1\hat{a}_2^{\dagger}]=2i(\hat{a}_1^{\dagger}\hat{a}_1
	-\hat{a}_2^{\dagger}\hat{a}_2)
	\end{array} \right. \\
	& & \left\{
	\begin{array}{ll}
	\! \! \! \! \! \! & [\hat{N}, \hat{a}_1^{\dagger}\hat{a}_1
	-\hat{a}_2^{\dagger}\hat{a}_2]=0 \\
	\! \! \! \! \! \! & [\hat{N}, \hat{a}_1^{\dagger}\hat{a}_2
	+\hat{a}_1\hat{a}_2^{\dagger}]=0
	\end{array} \right.
\end{eqnarray*}

If we define

\begin{equation}
	\hat{L}_{\pm}\equiv (\hat{a}_1^{\dagger}\hat{a}_2
	+\hat{a}_1\hat{a}_2^{\dagger})\pm i(\hat{a}_1^{\dagger}\hat{a}_1
	-\hat{a}_2^{\dagger}\hat{a}_2) \text{,}
\end{equation}

\noindent the following commutation relations are derived.

\begin{eqnarray}
	& & [\hat{L}, \hat{L}_{\pm}]=\pm 2\hat{L}_{\pm} \\
	& & [\hat{N}, \hat{L}_{\pm}]=0
\end{eqnarray}

Therefore, $\hat{L}_{\pm}$ raises the eigenvalue of $\hat{L}$ by $\pm 2$ and does not change that of $\hat{N}$. Furthermore, investigation of the norm of $\hat{L}_{\pm}$ leads us to determining the range of $l$.

\begin{eqnarray*}
	& & \left\{
	\begin{array}{l}
	\hat{L}_{+}\hat{L}_{-}=\hat{N}^2-\hat{L}^2+2\hat{N}+2\hat{L} \\
	\hat{L}_{-}\hat{L}_{+}=\hat{N}^2-\hat{L}^2+2\hat{N}-2\hat{L}
	\end{array} \right. \\
	& & \langle N,l|\hat{L}_{\pm}\hat{L}_{\mp}
	|N,l \rangle =(N\pm l)(N\mp l+2)\geq 0 \\
	& & \text{thus,} \quad -N\leq l\leq N \quad \text{or} \quad 
	l=-N,-(N-2),-(N-4),\dots ,N-4,N-2,N
\end{eqnarray*}

This is because we call $\hat{L}$ $\ll$angular momentum$\gg$.

If we take $\theta m^2=-1$, there exists on-shell state $l=-N$ for all $N$.

At last, since we have obtained the complete system, we can decompose the field $\phi$ into it's component fields in the following way.

\begin{eqnarray}
	\phi (t_1,t_2) 
	& \! \! \! = & \! \! \! \sum_{N=0}^{\infty} \sum_{-N\leq l\leq N} 
	\alpha_{N,l}\phi_{N,l}(t_1,t_2) \nonumber \\
	& \! \! \! = & \! \! \! \sum_{N=0}^{\infty} \sum_{-N\leq l\leq N} 
	\alpha_{N,l}\langle t_1,t_2|N,l\rangle
\end{eqnarray}

The action, which produces the equation of motion (13), is

\begin{equation*}
	S=\int dt_1dt_2 \Big[ \partial_i \phi^{\dagger} \partial_i \phi 
	+\frac{1}{4\theta^2} t_it_i \phi^{\dagger}\phi +\frac{i}{2\theta} 
	\big\{ \phi^{\dagger}\epsilon_{ij}t_i\partial_j\phi 
	-\phi\epsilon_{ij}t_i\partial_j\phi^{\dagger} \big\}
	+m^2\phi^{\dagger}\phi \Big] \text{.}
\end{equation*}

Here, to construct the action, we must incorporate the field $\phi^{\dagger}$ into the theory.

Now, in order to finish the discussion in this section, we argue the field $\phi^{\dagger}$. The method is very similar to that of $\phi$, so we only sketch the main results.

\begin{eqnarray}
	& & \phi^{\dagger} (t_1,t_2)=\sum_{N=0}^{\infty} \sum_{-N\leq l\leq N} 
	\alpha_{N,l}^{\dagger} \langle N,l|t_1,t_2\rangle \\
	\text{equation } \! \! \! \! \! \! & & \! \! \! \text{of motion} 
	\nonumber \\ 
	& & \Big[ -(\partial_1^2+\partial_2^2)+\frac{1}{4\theta^2}(t_1^2+t_2^2)
	-\frac{i}{\theta}(t_1\partial_2-t_2\partial_1)+m^2\Big] \phi^{\dagger}
	(t_1,t_2)=0 \\
	\text{mass-shell} \! \! \! \! & & \! \! \! \text{condition} 
	\nonumber \\
	& & \frac{1}{\theta}(N+1)-\frac{1}{\theta}l+m^2=0
\end{eqnarray}

This Klein-Gordon equation is made form the following noncommutative space and Einstein's energy-momentum relation : 

\begin{eqnarray*}
	& & [\hat{x}^{\prime}_i ,\hat{x}^{\prime}_j ] =-i\theta_{ij} \\
	& & [\hat{x}^{\prime}_i ,\hat{p}^{\prime}_j ] =i\hbar \delta_{ij} 
	\qquad 	(i,j=1\sim n) \\
	& & [\hat{p}^{\prime}_i ,\hat{p}^{\prime}_j ] =i\theta^{-1}_{ij} \\
	& & \left\{
	\begin{array}{l}
	\hat{x}^{\prime}_1=\frac{1}{2}t_1-i\theta \partial_2 \\
	\hat{x}^{\prime}_2=\frac{1}{2}t_2+i\theta \partial_1 
	\end{array} \right.
	, \quad 
	\left\{
	\begin{array}{l}
	\hat{p}^{\prime}_1=-\frac{1}{2\theta}t_2-i\partial_1 \\
	\hat{p}^{\prime}_2=\frac{1}{2\theta}t_1-i\partial_2 
	\end{array} \right. \\
	& & \\
	& & (\hat{p}^{\prime \; 2}_1+\hat{p}^{\prime \; 2}_2+m^2)\phi^{\dagger}
	(t_1,t_2)=0
\end{eqnarray*}

The operators $\hat{x}^{\prime}_i$s, $\hat{p}^{\prime}_i$s belong to $End_{\mathcal{A}} \mathcal{H}$, because these all commute with $\hat{x}_i$s and $\hat{p}_i$s. So, we can say the field $\phi^{\dagger}$ lives on noncommutative space $End_{\mathcal{A}} \mathcal{H}$. As is well known, the gauge field interacting with $\phi$ also belongs to $End_{\mathcal{A}} \mathcal{H}$.

Although the usual scalar field theory on commutative space has the continuous spectrum, we find that the above model has the discrete spectrum $(N,l)$. This is the desirable result for quantum mechanics on noncommutative space. Because of the uncertainty of the space-time itself (and also the momentum), the spectrum may gain the discreteness.

So far we have discussed the ``Euclidean" space, but if we want to consider on the ``Minkowskian" space such that $-\hat{p}_1^2+\hat{p}_2^2+m^2=0$, we can't first-quantize by using harmonic oscillator system. The ``Minkowskian" space will be discussed in future works.

\vskip 1.5cm

\noindent
{\bf 4. Second quantization : scalar field theory on noncommutative space}

\vskip 0.5cm

To begin this section, let us write down again the action which produces the equation of motion (13)(27),

\begin{equation}
	S=\int dt_1dt_2 \Big[ \partial_i \phi^{\dagger} \partial_i \phi 
	+\frac{1}{4\theta^2} t_it_i \phi^{\dagger}\phi +\frac{i}{2\theta} 
	\big\{ \phi^{\dagger}\epsilon_{ij}t_i\partial_j\phi 
	-\phi\epsilon_{ij}t_i\partial_j\phi^{\dagger} \big\}
	+m^2\phi^{\dagger}\phi \Big]
\end{equation}

\noindent with $\epsilon_{12}=1$. To second-quantize the fields $\phi$ and $\phi^{\dagger}$, we define its conjugate momenta $\pi_1$ and $\pi^{\dagger}_1$ with time-coordinate being $t_1$.

\begin{eqnarray}
	& & \left\{
	\begin{array}{l}
	\pi_1 \equiv \displaystyle \frac{\delta S}{\delta(\partial_1 \phi)} 
	=(\partial_1-\frac{i}{2\theta}t_2)\phi^{\dagger} \\
	\\
	\pi^{\dagger}_1 \equiv 
	\displaystyle \frac{\delta S}{\delta(\partial_1 \phi^{\dagger})} 
	=(\partial_1+\frac{i}{2\theta}t_2)\phi
	\end{array} \right.
\end{eqnarray}

For the moment, we shall closely look at the operators in (30). This is needed for getting the algebra of the coefficients $\alpha_{N,l},\; \alpha_{N,l}^{\dagger}$. The following commutation relations can be easily checked : 

\begin{eqnarray}
	& \left\{
	\begin{array}{ll}
	\! \! \! & [\hat{L}, (\partial_1-\displaystyle \frac{i}{2\theta}t_2)]
	=-i(\partial_2+\frac{i}{2\theta}t_1) \\
	\! \! \! & [\hat{L}, (\partial_2+\displaystyle \frac{i}{2\theta}t_1)]
	=i(\partial_1-\frac{i}{2\theta}t_2)
	\end{array} \right. , 
	& \left\{
	\begin{array}{ll}
	\! \! \! & [\hat{L}, (\partial_1+\displaystyle \frac{i}{2\theta}t_2)]
	=-i(\partial_2-\frac{i}{2\theta}t_1) \\
	\! \! \! & [\hat{L}, (\partial_2-\displaystyle \frac{i}{2\theta}t_1)]
	=i(\partial_1+\frac{i}{2\theta}t_2)
	\end{array} \right. \\
	& & \nonumber \\
	& \left\{
	\begin{array}{ll}
	\! \! \! & [\hat{N}, (\partial_1-\displaystyle \frac{i}{2\theta}t_2)]
	=i(\partial_2+\frac{i}{2\theta}t_1) \\
	\! \! \! & [\hat{N}, (\partial_2+\displaystyle \frac{i}{2\theta}t_1)]
	=-i(\partial_1-\frac{i}{2\theta}t_2)
	\end{array} \right. , 
	& \left\{
	\begin{array}{ll}
	\! \! \! & [\hat{N}, (\partial_1+\displaystyle \frac{i}{2\theta}t_2)]
	=-i(\partial_2-\frac{i}{2\theta}t_1) \\
	\! \! \! & [\hat{N}, (\partial_2-\displaystyle \frac{i}{2\theta}t_1)]
	=i(\partial_1+\frac{i}{2\theta}t_2)
	\end{array} \right.
\end{eqnarray}

Let us define the ``raising and lowering" operators : 

\begin{eqnarray}
	& & \hat{N}_{\pm}\equiv i\big[(\partial_1-\displaystyle 
	\frac{i}{2\theta}t_2)\pm i(\partial_2+\displaystyle 
	\frac{i}{2\theta}t_1)\big] \\
	& & \hat{N}_{\pm}^{\dagger}\equiv -i\big[(\partial_1+\displaystyle 
	\frac{i}{2\theta}t_2)\mp i(\partial_2-\frac{i}{2\theta}t_1)\big] 
	\text{.}
\end{eqnarray}

\begin{eqnarray}
	\left\{
	\begin{array}{ll}
	\! \! \! \! \! & [\hat{N}, \hat{N}_{\pm}]=\pm \hat{N}_{\pm} \\
	\! \! \! \! \! & [\hat{L}, \hat{N}_{\pm}]=\mp \hat{N}_{\pm} 
	\end{array} \right. ,\qquad 
	\left\{
	\begin{array}{ll}
	\! \! \! \! \! & [\hat{N}, \hat{N}_{\pm}^{\dagger}]
	=\pm \hat{N}_{\pm}^{\dagger} \\
	\! \! \! \! \! & [\hat{L}, \hat{N}_{\pm}^{\dagger}]
	=\pm \hat{N}_{\pm}^{\dagger} 
	\end{array} \right.
\end{eqnarray}

Consequently, $\hat{N}_{\pm}$ raises the eigenvalues of $\hat{N}$ and $\hat{L}$ by $\pm 1$ and $\mp 1$ respectively, and $\hat{N}_{\pm}^{\dagger}$ does so to $\hat{N}$, $\hat{L}$ by $\pm 1$, $\pm 1$. So, if $\phi$ and $\phi^{\dagger}$ are acted on by $\hat{N}_{\pm}$ and $\hat{N}_{\pm}^{\dagger}$ respactively, they remain in the on-shell states, in other words, keep the mass-shell conditions (21)(28).

Next, we shall investigate the norms of $\hat{N}_{\pm}$ and $\hat{N}_{\pm}^{\dagger}$. This can be done by the following equations : 

\begin{eqnarray}
	& & \left\{
	\begin{array}{l}
	\hat{N}_{\pm}\hat{N}_{\mp}=\frac{1}{\theta}(\hat{N}+1)
	-\frac{1}{\theta}\hat{L}\mp \frac{1}{\theta} \\
	\langle N,l|\hat{N}_{\pm}\hat{N}_{\mp}|N,l\rangle 
	=\frac{1}{\theta}\big\{ (N+1)-l\mp 1\big\} 
	\end{array} \right. \nonumber \\
	& & \hat{N}_{\pm}|N,l\rangle 
	=-\frac{1}{\sqrt{\theta}}\sqrt{(N+1)-l\pm 1}|N\pm 1,l\mp 1\rangle 
	\text{,} \\
	& & \left\{
	\begin{array}{l}
	\hat{N}_{\pm}^{\dagger}\hat{N}_{\mp}^{\dagger}=\frac{1}{\theta}
	(\hat{N}+1)+\frac{1}{\theta}\hat{L}\mp \frac{1}{\theta} \\
	\langle N,l|\hat{N}_{\pm}^{\dagger}\hat{N}_{\mp}^{\dagger}|N,l\rangle 
	=\frac{1}{\theta}\big\{ (N+1)+l\mp 1\big\} 
	\end{array} \right. \nonumber \\
	& & \hat{N}_{\pm}^{\dagger}|N,l\rangle 
	=-\frac{1}{\sqrt{\theta}}\sqrt{(N+1)+l\pm 1}|N\pm 1,l\pm 1\rangle 
	\text{.}
\end{eqnarray}

Now that we have understood the operators appearing in the momenta $\pi_1 ,\pi^{\dagger}_1$, let us quantize the fields $\phi ,\; \phi^{\dagger}$. The procedure is familiar one and gives the ``equal-time" commutation relations between the fields $\phi ,\; \phi^{\dagger}$ and its conjugate momenta $\pi_1 ,\pi^{\dagger}_1$.

\begin{eqnarray}
	\left\{
	\begin{array}{ll}
	\! \! \! \! \! & [\hat{\phi}(t_1,t_2), \hat{\pi}_1(t_1,t_2^{\prime})]
	=i\delta (t_2-t_2^{\prime}) \\
	\! \! \! \! \! & [\hat{\phi}^{\dagger}(t_1,t_2), \hat{\pi}^{\dagger}_1
	(t_1,t_2^{\prime})]
	=i\delta (t_2-t_2^{\prime}) \\
	\! \! \! \! \! & [\hat{\phi}(t_1,t_2), \hat{\phi}(t_1,t_2^{\prime})]
	=[\hat{\phi}^{\dagger}(t_1,t_2), \hat{\phi}^{\dagger}(t_1,t_2^{\prime})]
	=0 \\
	\! \! \! \! \! & [\hat{\pi}_1(t_1,t_2), \hat{\pi}_1(t_1,t_2^{\prime})]=
	[\hat{\pi}^{\dagger}_1(t_1,t_2), \hat{\pi}^{\dagger}_1
	(t_1,t_2^{\prime})]=0
	\end{array} \right.
\end{eqnarray}

These are all the Schr\"{o}dinger operators. 

The fields $\phi ,\phi^{\dagger}$ are expanded as 

\begin{eqnarray}
	& & \phi (t_1,t_2)=\sum_{N=0}^{\infty}\alpha_{N,-N-\theta m^2-1} 
	\langle t_1,t_2|N,-N-\theta m^2-1\rangle \\
	& & \phi^{\dagger}(t_1,t_2)=\sum_{N=0}^{\infty}
	\alpha^{\dagger}_{N,N+\theta m^2+1} \langle N,N+\theta m^2+1|
	t_1,t_2\rangle \text{,}
\end{eqnarray}

\noindent as the solution of (13),(27) respectively. Using these expansion, we have 

\begin{eqnarray}
	& & \Big[ \phi(t_1,t_2),\pi_1(t_1,t_2^{\prime})\Big]
	=i\delta (t_2-t_2^{\prime}) \nonumber \\
	& & \Big[ \sum_{N} \hat{\alpha}_{N,-N-\theta m^2-1} \langle t_1,t_2 |
	N,-N-\theta m^2-1\rangle , \nonumber \\
	& & \qquad \big( -\frac{i}{2}\big) \sum_{N^{\prime \prime}}
	\hat{\alpha}^{\dagger}
	_{N^{\prime \prime},N^{\prime \prime}+\theta m^2+1}
	\langle N^{\prime \prime},N^{\prime \prime}+\theta m^2+1|
	(\hat{N}_{+}+\hat{N}_{-})|t_1,t_2^{\prime}\rangle \Big]=i\delta 
	(t_2-t_2^{\prime}) \nonumber \\
	& & \Big[ \sum_{N} \hat{\alpha}_{N,-N-\theta m^2-1} \langle t_1,t_2 |
	N,-N-\theta m^2-1\rangle , \nonumber \\
	& & \qquad \big( \frac{i}{2\sqrt{\theta}}\big) \sum_{N^{\prime \prime}}
	\hat{\alpha}^{\dagger}
	_{N^{\prime \prime},N^{\prime \prime}+\theta m^2+1}
	\big\{ \sqrt{-\theta m^2+1} \langle N^{\prime \prime}+1
	,N^{\prime \prime}+\theta m^2|t_1,t_2^{\prime}\rangle \nonumber \\
	& & \qquad \qquad \qquad \quad +\sqrt{-\theta m^2-1} \langle 
	N^{\prime \prime}-1,N^{\prime \prime}+\theta m^2+2|t_1,t_2^{\prime}
	\rangle \big\} \Big] =i\delta (t_2-t_2^{\prime}) \text{,} \\
	& & \text{ } \nonumber \\
	& & \Big[ \phi^{\dagger}(t_1,t_2),\pi^{\dagger}_1(t_1,t_2^{\prime})
	\Big] =i\delta (t_2-t_2^{\prime}) \nonumber \\
	& & \Big[ \sum_{N^{\prime}} 
	\hat{\alpha}^{\dagger}_{N^{\prime},N^{\prime}+\theta m^2+1} \langle 
	N^{\prime},N^{\prime}+\theta m^2+1 |t_1,t_2\rangle , \nonumber \\
	& & \qquad \big( \frac{i}{2}\big) \sum_{N^{\prime \prime}}
	\hat{\alpha}_{N^{\prime \prime},-N^{\prime \prime}-\theta m^2-1}
	\langle t_1,t_2^{\prime}|(\hat{N}_{+}+\hat{N}_{-})|
	N^{\prime \prime},N^{\prime \prime}+\theta m^2+1\rangle \Big]=i\delta 
	(t_2-t_2^{\prime}) \nonumber \\
	& & \Big[ \sum_{N^{\prime}} 
	\hat{\alpha}^{\dagger}_{N^{\prime},N^{\prime}+\theta m^2+1} \langle 
	N^{\prime},N^{\prime}+\theta m^2+1|t_1,t_2\rangle , \nonumber \\
	& & \qquad \big( \frac{i}{2\sqrt{\theta}}\big) \sum_{N^{\prime \prime}}
	\hat{\alpha}_{N^{\prime \prime},-N^{\prime \prime}-\theta m^2-1}
	\big\{ -\sqrt{-\theta m^2+1} \langle t_1,t_2^{\prime}|
	N^{\prime \prime}+1,-N^{\prime \prime}-\theta m^2\rangle \nonumber \\
	& & \qquad \qquad \qquad \quad -\sqrt{-\theta m^2-1} \langle 
	t_1,t_2^{\prime}|N^{\prime \prime}-1,-N^{\prime \prime}-\theta m^2-2
	\rangle \big\} \Big] =i\delta (t_2-t_2^{\prime}) \text{.}
\end{eqnarray}

By acting the following integration operators on each equations such that 

\begin{eqnarray*}
	& & \int dt_2\int dt_2^{\prime}\langle N,l|t_1,t_2\rangle 
	\langle t_1,t_2^{\prime}|N^{\prime},l^{\prime}\rangle \quad 
	\text{acts on }(41) \text{,} \\
	& & \int dt_2\int dt_2^{\prime}\langle t_1,t_2|N^{\prime},l^{\prime}
	\rangle \langle N,l|t_1,t_2^{\prime}\rangle \quad 
	\text{acts on }(42) \text{,}
\end{eqnarray*}

\noindent and using the completeness conditions : $\displaystyle \int dt_2 |t_1,t_2\rangle \langle t_1,t_2|=1$, we can obtain the commutation relations between the coefficients $\alpha_{N,l}, \; \alpha^{\dagger}_{N,l}$.

From (41), 

\begin{eqnarray}
	\left.
	\begin{array}{l}
	\Big[ \hat{\alpha}_{N,-N-\theta m^2-1}, \big( \frac{1}{2} 
	\sqrt{-m^2+\textstyle \frac{1}{\theta}}\big) \hat{\alpha}^{\dagger}
	_{N^{\prime}-1,N^{\prime}+\theta m^2}\Big] =\delta_{NN^{\prime}} 
	\delta_{-N-\theta m^2-1,N^{\prime}+\theta m^2-1} \\
	\\
	\Big[ \hat{\alpha}_{N,-N-\theta m^2-1}, \big( \frac{1}{2} 
	\sqrt{-m^2-\textstyle \frac{1}{\theta}}\big) \hat{\alpha}^{\dagger}
	_{N^{\prime}+1,N^{\prime}+\theta m^2+2}\Big] =\delta_{NN^{\prime}} 
	\delta_{-N-\theta m^2-1,N^{\prime}+\theta m^2+3} \text{,}
	\end{array} \right.
\end{eqnarray}

\noindent and from (42), 

\begin{eqnarray}
	\left.
	\begin{array}{l}
	\Big[ \hat{\alpha}^{\dagger}_{N^{\prime},N^{\prime}+\theta m^2+1}, 
	\big( -\frac{1}{2} \sqrt{-m^2+\textstyle \frac{1}{\theta}}\big) 
	\hat{\alpha}_{N-1,-N-\theta m^2}\Big] =\delta_{NN^{\prime}} 
	\delta_{-N-\theta m^2+1,N^{\prime}+\theta m^2+1} \\
	\\
	\Big[ \hat{\alpha}^{\dagger}_{N^{\prime},N^{\prime}+\theta m^2+1}, 
	\big( -\frac{1}{2} \sqrt{-m^2-\textstyle \frac{1}{\theta}}\big) 
	\hat{\alpha}_{N+1,-N-\theta m^2-2}\Big] =\delta_{NN^{\prime}} 
	\delta_{-N-\theta m^2-3,N^{\prime}+\theta m^2+1} \text{,}
	\end{array} \right.
\end{eqnarray}

Thus, we have the full algebra.

\begin{eqnarray}
	& & \Big[ \hat{\alpha}_{-\theta m^2,-1}, 
	\hat{\alpha}^{\dagger}_{-\theta m^2-1,0} \Big]
	=\frac{2}{\sqrt{-m^2+\frac{1}{\theta}}} \\
	& & \Big[ \hat{\alpha}_{-\theta m^2-2,1}, 
	\hat{\alpha}^{\dagger}_{-\theta m^2-1,0} \Big]
	=\frac{2}{\sqrt{-m^2-\frac{1}{\theta}}} \\
	& & \Big[ \hat{\alpha}_{-\theta m^2-1,0}, 
	\hat{\alpha}^{\dagger}_{-\theta m^2,1} \Big]
	=\frac{2}{\sqrt{-m^2+\frac{1}{\theta}}} \\
	& & \Big[ \hat{\alpha}_{-\theta m^2-1,0}, 
	\hat{\alpha}^{\dagger}_{-\theta m^2-2,-1} \Big]
	=\frac{2}{\sqrt{-m^2-\frac{1}{\theta}}} \\
	& & \text{All others are commuting.} \nonumber
\end{eqnarray}

From this algebra, we see that the amplitude $\langle \phi \phi^{\dagger} \rangle$ has only 4-excitation states and this result is unpleasant. This is because $\phi$'s mass-shell condition is different from $\phi^{\dagger}$'s. So I think that we must impose noncommutativity between them as

\begin{eqnarray*}
	& & [\phi (t_1,t_2),\phi (t_1,t_2^{\prime})]=i\theta(t_2-t_2^{\prime}) 
	\\
	& & [\phi^{\dagger} (t_1,t_2),\phi^{\dagger} (t_1,t_2^{\prime})]
	=-i\theta(t_2-t_2^{\prime}) \qquad \text{etc.} \\
	& & \theta (x)=
	\left\{ \begin{array}{ll}
	+1 & ,x>0 \\
	0 & ,x=0 \\
	-1 & ,x<0
	\end{array} \right. \text{.}
\end{eqnarray*}

This ``deformed" second-quantization will be discussed in future's works.

\vskip 1.5cm

\noindent
{\bf 5. Second quantization : functional integral}

\vskip 0.5cm

In this section, we will construct the functional integral formalism of the second quantization of the scalar field theory on noncommutative space. According to Ref.~\cite{zinn}, we start with the ``Minkowskian" version of the action $(29)$, and use the ``imaginary-time" evolution operator in order to define the ``Euclidean" functional integral. Although our method is very similar to that of usual scalar field theories, I would like to follow the calculation carefully, because our result is slightly different.

The ``Minkowskian" Klein-Gordon model is defined in the following way.

\begin{eqnarray}
	& & \left\{
	\begin{array}{l}
	\hat{x}_1=\frac{1}{2}t_1+i\theta \partial_2 \\
	\hat{x}_2=\frac{1}{2}t_2-i\theta \partial_1 
	\end{array} \right.
	, \quad 
	\left\{
	\begin{array}{l}
	\hat{p}_1=-\frac{1}{2\theta}t_2+i\partial_1 \\
	\hat{p}_2=-\frac{1}{2\theta}t_1-i\partial_2 
	\end{array} \right. \\
	& & \nonumber \\
	& & (-\hat{p}_1^2+\hat{p}_2^2+m^2)\phi(t_1,t_2)=0 \\
	& & \Big[ (\partial_1^2-\partial_2^2)+\frac{1}{4\theta^2}(t_1^2-t_2^2)
	+\frac{i}{\theta}(t_1\partial_2+t_2\partial_1)+m^2\Big] \phi(t_1,t_2)=0
	\\
	S_M \! \! \! \! \! \! \! \! & & =\int dt_1dt_2 \Big[ \partial_1 
	\phi^{\dagger} \partial_1 \phi -\partial_2 \phi^{\dagger} \partial_2 
	\phi -\frac{1}{4\theta^2}(t_1^2-t_2^2)\phi^{\dagger}\phi \nonumber \\
	& & \qquad \qquad \qquad \quad -\frac{i}{2\theta}(\phi^{\dagger}t_1
	\partial_2\phi +\phi^{\dagger}t_2\partial_1\phi -\phi t_1
	\partial_2\phi^{\dagger}-\phi t_2\partial_1\phi^{\dagger})
	-m^2\phi^{\dagger}\phi \Big] \\
	& & \equiv \int dt_1dt_2 \mathfrak{L}_M \nonumber
\end{eqnarray}

\begin{eqnarray}
	\left\{
	\begin{array}{l}
	\pi_1 \equiv \displaystyle \frac{\delta S}{\delta(\partial_1 \phi)} 
	=(\partial_1-\frac{i}{2\theta}t_2)\phi^{\dagger} \\
	\\
	\pi^{\dagger}_1 \equiv 
	\displaystyle \frac{\delta S}{\delta(\partial_1 \phi^{\dagger})} 
	=(\partial_1+\frac{i}{2\theta}t_2)\phi
	\end{array} \right.
\end{eqnarray}

The definition of the Hamiltonian is supposed to be the same as the usual one.

\begin{eqnarray}
	& \mathfrak{H} \! \! \! & \equiv \pi_1 \partial_1 \phi 
	+\pi_1^{\dagger} \partial_1 \phi^{\dagger} -\mathfrak{L}_M \nonumber \\
	& & =\partial_1 \phi^{\dagger} \partial_1 \phi +
	\partial_2 \phi^{\dagger} \partial_2 \phi +\frac{1}{4\theta^2}
	(t_1^2-t_2^2)\phi^{\dagger}\phi +\frac{i}{2\theta}(\phi^{\dagger}t_1
	\partial_2\phi -\phi t_1\partial_2\phi^{\dagger})
	+m^2\phi^{\dagger}\phi \nonumber \\
	& \! \! \! & =(\pi_1^{\dagger}-\frac{i}{2\theta}t_2 \phi)(\pi_1
	+\frac{i}{2\theta}t_2 \phi^{\dagger})
	+\partial_2 \phi^{\dagger} \partial_2 \phi +\frac{1}{4\theta^2}
	(t_1^2-t_2^2)\phi^{\dagger}\phi \nonumber \\
	& \! \! \! & \qquad \qquad \qquad \qquad \qquad \qquad \qquad \qquad 
	+\frac{i}{2\theta}(\phi^{\dagger}t_1
	\partial_2\phi -\phi t_1\partial_2\phi^{\dagger})
	+m^2\phi^{\dagger}\phi \\
	H(t_1) \! \! \! & \equiv \! \! \! & \int dt_2 \mathfrak{H} \equiv \int 
	dt_2 (\pi_1^{\dagger}-\frac{i}{2\theta}t_2 \phi)(\pi_1
	+\frac{i}{2\theta}t_2 \phi^{\dagger})+V[\phi ,\phi^{\dagger};t_1]
\end{eqnarray}

Then, we shall impose the canonical commutation relations.

\begin{eqnarray}
	& & \left\{
	\begin{array}{ll}
	\! \! \! \! \! & [\hat{\phi}(t_1,t_2), \hat{\pi}_1(t_1,t_2^{\prime})]
	=i\delta (t_2-t_2^{\prime}) \\
	\! \! \! \! \! & [\hat{\phi}^{\dagger}(t_1,t_2), \hat{\pi}_1^{\dagger}
	(t_1,t_2^{\prime})]=i\delta (t_2-t_2^{\prime})
	\end{array} \right. \\
	& & \left\{
	\begin{array}{ll}
	\! \! \! \! \! & \! \! \! \! \! \hat{\pi}_1(t_1,t_2)
	=-i \displaystyle \frac{\delta}{\delta \phi (t_1,t_2)} \\
	& \\
	\! \! \! \! \! & \! \! \! \! \! \hat{\pi}_1^{\dagger}(t_1,t_2)
	=-i \displaystyle \frac{\delta}{\delta \phi^{\dagger} (t_1,t_2)}
	\end{array} \right.
	\quad \text{ : Schr\"{o}dinger operator}
\end{eqnarray}

Suppose that $\hat{\phi}_H(t_1,t_2)$ and $\hat{\phi}_H^{\dagger}(t_1,t_2)$ are Heisenberg operators at ``time" $t_1$. In Heisenberg representation, we have

\begin{eqnarray*}
	& & \hat{\phi}_H(t_1,t_2) |\phi ;t_1\rangle_H =\phi |\phi ;t_1
	\rangle_H \\
	& & \hat{\phi}_H^{\dagger}(t_1,t_2) |\phi ;t_1\rangle_H 
	=\phi^{\dagger} |\phi ;t_1\rangle_H \\
	& & \text{completeness} \quad \int \mathfrak{D}\phi \mathfrak{D} 
	\phi^{\dagger} |\phi ;t_1\rangle_H {}_H\langle \phi ;t_1|=1 \\
	& & \text{orthonormality} \quad {}_H\langle \phi ;t_1|
	\phi^{\prime} ;t_1\rangle_H =\delta^2 [\phi -\phi^{\prime}] \text{.}
\end{eqnarray*}

Then, let us investigate the transition amplitude from ``time" $t_{1,i}$ to $t_{1,f}$, and cut the interval $(t_{1,f}-t_{1,i})$ into a small piece $\epsilon \ll 1$, where $(t_{1,f}-t_{1,i})=(N+1)\epsilon$, $N\gg 1$.

\begin{eqnarray}
	& & {}_H\langle \phi_f ;t_{1,f}|\phi_i ;t_{1,i}\rangle_H \nonumber \\
	& & \qquad =\int \mathfrak{D}\phi_N\mathfrak{D}\phi_N^{\dagger} \dots 
	\int \mathfrak{D}\phi_1\mathfrak{D}\phi_1^{\dagger} \nonumber \\
	& & \qquad \qquad \times {}_H\langle \phi_f ;t_{1,f}|\phi_N ;t_{1,N}
	\rangle_H{}_H\langle \phi_N ;t_{1,N}|\dots |\phi_1 ;t_{1,1}\rangle_H
	{}_H\langle \phi_1 ;t_{1,1}|\phi_i ;t_{1,i}\rangle_H \\
	& & \text{Here, } \quad \mathfrak{D}\phi_k \equiv 
	\prod_{t_{2,k}=-\infty}^{\infty} \mathfrak{D}\phi_k (t_{1,k},t_{2,k}) 
	\nonumber
\end{eqnarray}

Now, we shall go to the Schr\"{o}dinger representation. Let us express the evolution operator as $\hat{U}(t_1,t_1^{\prime})$. We obtain

\begin{eqnarray}
	\text{Schr\"{o}dinger eq.} & & -\partial_1 \hat{U}(t_1,t_1^{\prime})
	=\hat{H}(t_1) \hat{U}(t_1,t_1^{\prime}) \\
	& & \qquad \text{for } t_1\geq t_1^{\prime} ,\quad \hat{U}(t_1,t_1)=1 
	\text{.} \nonumber
\end{eqnarray}

Since the Hamiltonian manifestly depends on the ``time" $t_1$, $\hat{U}(t_1,t_1^{\prime}) \neq e^{-\hat{H}(t_1-t_1^{\prime})}$.

Next, we must evaluate the transition amplitude during the small interval $(t_{1,k}-t_{1,k-1})=\epsilon$ by using the Schr\"{o}dinger representation $|\phi \rangle_S$. 
\begin{eqnarray}
	{}_H\langle \phi_k ;t_{1,k}|\phi_{k-1} ;t_{1,k-1}\rangle_H 
	={}_S\langle \phi_k |\hat{U}(t_{1,k},t_{1,k-1})|\phi_{k-1} \rangle_S
\end{eqnarray}

Substituting $(60)$ into the Schr\"{o}dinger equation $(59)$, we have

\begin{eqnarray}
	& & -\partial_{1,k} \; {}_S\langle \phi_k |\hat{U}(t_{1,k},t_{1,k-1})
	|\phi_{k-1} \rangle_S \nonumber \\
	& & \qquad \qquad =\hat{H}(t_{1,k})\; {}_S\langle \phi_k |
	\hat{U}(t_{1,k},t_{1,k-1})|\phi_{k-1} \rangle_S \text{.}
\end{eqnarray}

Here, in replacing the classical Hamiltonian $H(t_1)$ by the quantum Hamiltonian $\hat{H}(t_1)$, the ambiguity of the operator ordering arises in the terms such as $\pi_1\phi,\; \pi_1^{\dagger}\phi^{\dagger}$. So, we define the quantum Hamiltonian by using the Weyl ordering. For example, 

\begin{eqnarray}
	& & \int_{-\infty}^{\infty} dt_2 \pi_1\phi \nonumber \\
	& & \rightarrow \int_{-\infty}^{\infty} dt_2 \big\{
	\hat{\pi}_1\hat{\phi} \big\}_W \nonumber \\
	& & \quad \equiv \frac{1}{2} \int_{-\infty}^{\infty} dt_2 \big\{ 
	\hat{\pi}_{1,k}(t_{1,k},t_2)\hat{\phi}_{k-1}(t_{1,k-1},t_2)
	+\hat{\phi}_k(t_{1,k},t_2)\hat{\pi}_{1,k-1}(t_{1,k-1},t_2) \big\} 
	\text{.}
\end{eqnarray}

Then,

\begin{eqnarray}
	& & \int_{-\infty}^{\infty} dt_2 \big\{\hat{\pi}_1\hat{\phi} 
	\big\}_W {}_S\langle \phi_k |\hat{U}(t_{1,k},t_{1,k-1})|\phi_{k-1} 
	\rangle_S \nonumber \\
	& & \quad =\frac{1}{2} \int_{-\infty}^{\infty} dt_2 \big\{ 
	-i\frac{\delta}{\delta \phi_k (t_{1,k},t_2)}\phi_{k-1} (t_{1,k-1},t_2)
	+\phi_k (t_{1,k},t_2)\cdot i\frac{\delta}{\delta \phi_{k-1} (t_{1,k-1}
	,t_2)} \big\} \nonumber\\
	& & \qquad \qquad \qquad \times {}_S\langle \phi_k |
	\hat{U}(t_{1,k},t_{1,k-1})|\phi_{k-1} \rangle_S \text{.}
\end{eqnarray}

The same thing can be said about $\pi_1^{\dagger}\phi^{\dagger}$. To solve the Schr\"{o}dinger equation $(61)$, let us define

\begin{eqnarray}
	& & \text{Fourier decomposition : }\nonumber \\
	& & {}_S\langle \phi_k |\hat{U}(t_{1,k},t_{1,k-1})|\phi_{k-1} 
	\rangle_S \nonumber \\
	& & \qquad \equiv \int \mathfrak{D}\pi_{1,k}\mathfrak{D}
	\pi_{1,k}^{\dagger} \; u_{\pi_{1,k}}(\epsilon ) \nonumber \\
	& & \qquad \qquad \times \exp \big[ i\int_{-\infty}^{\infty} dt_2 
	\pi_{1,k}(t_{1,k},t_2) \Big\{ \phi_k (t_{1,k},t_2)-\phi_{k-1} 
	(t_{1,k-1},t_2)\Big\} \nonumber \\
	& & \qquad \qquad \qquad \quad +i\int_{-\infty}^{\infty} dt_2 
	\pi_{1,k}^{\dagger}(t_{1,k},t_2) \Big\{ \phi_k^{\dagger} (t_{1,k},t_2)
	-\phi_{k-1}^{\dagger}(t_{1,k-1},t_2)\Big\} \big] \text{.} \\
	& & \qquad \text{boundary condition : } \; u_{\pi_{1,k}}(0)=1 \nonumber
\end{eqnarray}

Substituting $(64)$ into $(63)$ with omitting the $(t_1,t_2)$-dependence, we obtain

\begin{eqnarray*}
	& & \big\{ \hat{\pi}_1\hat{\phi}\big\}_W {}_S\langle \phi_k|\hat{U}
	(\epsilon )|\phi_{k-1}\rangle_S \\
	& & \quad =\Big\{ \big( -i\frac{\delta}{\delta \phi_k}\big) \phi_{k-1}
	+\phi_k\big( i\frac{\delta}{\delta \phi_{k-1}}\big) \Big\} \\
	& & \qquad \qquad \times \int \mathfrak{D}\pi_{1,k}\mathfrak{D}
	\pi_{1,k}^{\dagger} \; u_{\pi_{1,k}}(\epsilon ) \exp \big[ i\pi_{1,k}
	(\phi_k-\phi_{k-1})+i\pi_{1,k}^{\dagger}(\phi_k^{\dagger}
	-\phi_{k-1}^{\dagger}) \big] \\
	& & \quad =\int \mathfrak{D}\pi_{1,k}\mathfrak{D}\pi_{1,k}^{\dagger}
	\; \pi_{1,k}\frac{\phi_k+\phi_{k-1}}{2} u_{\pi_{1,k}}(\epsilon ) \exp 
	\big[ i\pi_{1,k}(\phi_k-\phi_{k-1})+i\pi_{1,k}^{\dagger}
	(\phi_k^{\dagger}-\phi_{k-1}^{\dagger}) \big] \\
	& & \text{The same for } \pi_1^{\dagger}\phi^{\dagger} \text{.}
\end{eqnarray*}

At last, let us solve the Sch\"{o}dinger equation $(61)$. The equation is

\begin{eqnarray*}
	& & -\partial_{\epsilon}\; {}_S\langle \phi_k|\hat{U}(\epsilon )
	|\phi_{k-1}\rangle_S \\
	& & \qquad =\hat{H}(t_{1,k}){}_S\langle \phi_k|\hat{U}
	(\epsilon )|\phi_{k-1}\rangle_S \\
	& & \qquad =\big[ \Big\{ (\hat{\pi}_1^{\dagger}-\frac{i}{2\theta}t_2
	\hat{\phi})(\hat{\pi}_1+\frac{i}{2\theta}t_2\hat{\phi}^{\dagger})
	\Big\}_W +V[\hat{\phi},\hat{\phi}^{\dagger};t_{1,k}] \big]{}_S\langle 
	\phi_k|\hat{U}(\epsilon )|\phi_{k-1}\rangle_S \text{.} \\
	& & \\
	& & \! \! \! \! \! \! \! \! \! \! \text{For } u_{\pi_{1,k}}(\epsilon ) 
	\text{,} \\
	& & \\
	& & -\partial_{\epsilon}\; u_{\pi_{1,k}}(\epsilon ) \\
	& & \qquad =\Big\{ \pi_{1,k}^{\dagger}\pi_{1,k}-\frac{i}{2\theta}t_2
	\frac{\phi_k+\phi_{k-1}}{2}\pi_{1,k}+\frac{i}{2\theta}t_2
	\frac{\phi_k^{\dagger}+\phi_{k-1}^{\dagger}}{2}\pi_{1,k}^{\dagger} 
	+\frac{1}{4\theta^2}t_2^2\phi_k^{\dagger}\phi_k \\
	& & \qquad \qquad \qquad \qquad +V[\phi,\phi^{\dagger};t_{1,k-1}]+
	\frac{\partial V[\phi_k,\phi_k^{\dagger};t_{1,k}]}{\partial t_{1,k}}
	\Bigg|_{t_{1,k}=t_{1,k-1}} \epsilon +\cdot \cdot \cdot \Big\} 
	u_{\pi_{1,k}}(\epsilon ) \text{.}
\end{eqnarray*}

Thus, the solution is

\begin{eqnarray*}
	& & u_{\pi_{1,k}}(\epsilon ) =\exp \big[ -\epsilon 
	\Big\{ \pi_{1,k}^{\dagger}\pi_{1,k}-\frac{i}{2\theta}t_2
	\frac{\phi_k+\phi_{k-1}}{2}\pi_{1,k}+\frac{i}{2\theta}t_2
	\frac{\phi_k^{\dagger}+\phi_{k-1}^{\dagger}}{2}\pi_{1,k}^{\dagger} \\
	& & \qquad \qquad \qquad \qquad \qquad \qquad \qquad \qquad \qquad 
	+\frac{1}{4\theta^2}t_2^2\phi_k^{\dagger}\phi_k 
	+V[\phi,\phi^{\dagger};t_{1,k-1}] \Big\} +O(\epsilon^2) \big] \text{.}
\end{eqnarray*}

Therefore,

\begin{eqnarray*}
	& & {}_S\langle \phi_k|\hat{U}(\epsilon )|\phi_{k-1}\rangle_S \\
	& & \quad =\int \mathfrak{D}\pi_{1,k}\mathfrak{D}\pi_{1,k}^{\dagger}
	\exp \big[ i\pi_{1,k}(\phi_k-\phi_{k-1})+i\pi_{1,k}^{\dagger}
	(\phi_k^{\dagger}-\phi_{k-1}^{\dagger}) \\
	& & \quad \qquad -\epsilon \Big\{ 
	\pi_{1,k}^{\dagger}\pi_{1,k}-\frac{i}{2\theta}t_2
	\frac{\phi_k+\phi_{k-1}}{2}\pi_{1,k}+\frac{i}{2\theta}t_2
	\frac{\phi_k^{\dagger}+\phi_{k-1}^{\dagger}}{2}\pi_{1,k}^{\dagger} 
	+\frac{1}{4\theta^2}t_2^2\phi_k^{\dagger}\phi_k \\
	& & \qquad \qquad \qquad \qquad \qquad \qquad \qquad \qquad 
	+V[\phi,\phi^{\dagger};t_{1,k}] \Big\} +O(\epsilon^2) \big] \\
	& & \qquad \qquad \quad \text{Arrange the exponent in order of 
	power of } \pi_{1,k} \text{,} \\
	& & \quad =\int \mathfrak{D}\pi_{1,k}\mathfrak{D}\pi_{1,k}^{\dagger}
	\exp \big[ -\epsilon \pi_{1,k}^{\dagger}\pi_{1,k} \\
	& & \quad \qquad +i\Big\{ (\phi_k-\phi_{k-1})
	+\frac{\epsilon}{2\theta}t_2\frac{\phi_k+\phi_{k-1}}{2}\Big\} \pi_{1,k}
	+i\Big\{ (\phi_k^{\dagger}-\phi_{k-1}^{\dagger}) 
	-\frac{\epsilon}{2\theta}t_2\frac{\phi_k^{\dagger}
	+\phi_{k-1}^{\dagger}}{2}\Big\} \pi_{1,k}^{\dagger} \\
	& & \qquad \qquad \qquad \qquad \qquad \qquad \qquad \qquad 
	-\epsilon \Big\{ \frac{1}{4\theta^2}t_2^2\phi_k^{\dagger}\phi_k
	+V[\phi,\phi^{\dagger};t_{1,k}] \Big\} +O(\epsilon^2) \big] \\
	& & \qquad \qquad \quad \text{using the integration formula : } \\
	& & \qquad \qquad \qquad \left. \begin{array}{l}
	\int dzd\bar{z} \exp \big\{ -a\bar{z}z+ibz+i\bar{b}\bar{z} \big\} \\
	\sim \int dxdy \exp \big\{ -a(x^2+y^2)+i(b+\bar{b})x-(b-\bar{b})y 
	\big\} \\
	\sim \int dxdy \exp \big\{ -a(x-i\frac{b+\bar{b}}{2a})^2-a(y+
	\frac{b-\bar{b}}{2a})^2-\frac{(b+\bar{b})^2}{4a}
	+\frac{(b-\bar{b})^2}{4a} \big\} \\
	\sim \exp \big\{ -\frac{(b+\bar{b})^2}{4a}+\frac{(b-\bar{b})^2}{4a} 
	\big\} \\
	\sim \exp \big( -\frac{\bar{b}b}{a} \big)
	\end{array} \right. \\
	& & \quad =\exp \big[ -\frac{1}{\epsilon}\Big\{ (\phi_k^{\dagger}
	-\phi_{k-1}^{\dagger})-\frac{\epsilon}{2\theta}t_2
	\frac{\phi_k^{\dagger}+\phi_{k-1}^{\dagger}}{2}\Big\} \Big\{ 
	(\phi_k-\phi_{k-1})+\frac{\epsilon}{2\theta}t_2\frac{\phi_k
	+\phi_{k-1}}{2}\Big\} \\
	& & \qquad \qquad \qquad \qquad \qquad \qquad \qquad \qquad 
	-\epsilon \Big\{ \frac{1}{4\theta^2}t_2^2\phi_k^{\dagger}\phi_k
	+V[\phi,\phi^{\dagger};t_{1,k}]\Big\}+O(\epsilon^2) \big] \\
	& & \quad =\exp \big[ -\epsilon \Big\{ \big( \frac{\phi_k^{\dagger}
	-\phi_{k-1}^{\dagger}}{\epsilon}\big) \big( \frac{\phi_k-\phi_{k-1}}
	{\epsilon}\big) \\
	& & \qquad \qquad \qquad -\frac{1}{2\theta}t_2\frac{\phi_k^{\dagger}
	+\phi_{k-1}^{\dagger}}{2}\frac{\phi_k-\phi_{k-1}}{\epsilon}
	+\frac{1}{2\theta}t_2\frac{\phi_k+\phi_{k-1}}{2}\frac{\phi_k^{\dagger}
	-\phi_{k-1}^{\dagger}}{\epsilon} \\
	& & \qquad \qquad \qquad 
	+\frac{1}{4\theta^2}t_2^2\big( \phi_k^{\dagger}\phi_k 
	-\frac{\phi_k^{\dagger}+\phi_{k-1}^{\dagger}}{2} 
	\frac{\phi_k+\phi_{k-1}}{2}\big)
	+V[\phi,\phi^{\dagger};t_{1,k}]\Big\}+O(\epsilon^2)\big] \\
	& & \quad \equiv \exp \big[ -\epsilon \Big\{ *** \Big\} +O(\epsilon^2)
	\big] \text{.}
\end{eqnarray*}

We have omitted the overall factor. Now that we have found the transiton amplitude during the small interval $\epsilon$, we can get the full amplitude : 

\begin{eqnarray}
	& & {}_H\langle \phi_f ;t_{1,f}|\phi_i ;t_{1,i}\rangle_H \nonumber \\
	& & \qquad =\lim_{\genfrac{}{}{0pt}{1}{\epsilon \to 0}{N \to \infty}} 
	\prod_{n=1}^{N}\int \mathfrak{D}\phi_n \mathfrak{D}\phi_n^{\dagger} 
	\prod_{k=1}^{N+1} \exp \big[ -\epsilon \Big\{ *** \Big\} \big] 
	\nonumber \\
	& & \qquad =\lim_{\genfrac{}{}{0pt}{1}{\epsilon \to 0}{N \to \infty}} 
	\prod_{n=1}^{N}\int \mathfrak{D}\phi_n \mathfrak{D}\phi_n^{\dagger} 
	\exp \big[ -\epsilon \sum_{k=1}^{N+1} \Big\{ *** \Big\} \big] \nonumber
	 \\
	& & \qquad =\int \mathfrak{D}\phi \mathfrak{D}\phi^{\dagger} 
	\exp \big[ -\int_{t_{1,i}}^{t_{1,f}}dt_1 \int_{-\infty}^{\infty}dt_2 
	\Big\{ \partial_1\phi^{\dagger}\partial_1\phi -\frac{1}{2\theta}
	(\phi^{\dagger}t_2\partial_1\phi -\phi t_2\partial_1\phi^{\dagger}) 
	\nonumber \\
	& & \qquad \qquad +\partial_2\phi^{\dagger}\partial_2\phi 
	+\frac{1}{4\theta^2}(t_1^2-t_2^2)\phi^{\dagger}\phi +\frac{i}{2\theta}
	(\phi^{\dagger}t_1\partial_2\phi-\phi t_1\partial_2\phi^{\dagger})
	+m^2\phi^{\dagger}\phi \Big\} \big] \nonumber \\
	& & \qquad =\int \mathfrak{D}\phi \mathfrak{D}\phi^{\dagger} 
	\exp \big[ -\int_{t_{1,i}}^{t_{1,f}}dt_1 \int_{-\infty}^{\infty}dt_2 
	\Big\{ \partial_1\phi^{\dagger}\partial_1\phi +\partial_2
	\phi^{\dagger}\partial_2\phi \nonumber \\
	& & \qquad \qquad 
	+\frac{1}{4\theta^2}(t_1^2+t_2^2)\phi^{\dagger}\phi +\frac{i}{2\theta}
	(\phi^{\dagger}t_1\partial_2\phi-\phi^{\dagger}t_2\partial_1\phi -\phi 
	t_1\partial_2\phi^{\dagger}+\phi t_2\partial_1\phi^{\dagger})
	+m^2\phi^{\dagger}\phi \nonumber \\
	& & \qquad \qquad -\frac{1}{2\theta^2}t_2^2\phi^{\dagger}\phi 
	+(1+i)\frac{i}{2\theta}(\phi^{\dagger}t_2\partial_1\phi 
	-\phi t_2\partial_1\phi^{\dagger}) \Big\} \big] \text{.}
\end{eqnarray}

We see from this integration that, if we take the commutative limit $\theta \to 0$, we obtain the ordinary ``Euclidean" functional integral of scalar field theory on commutative space.

This is the functional integral of the (free) scalar field theory on noncommutative space. Note that the exponential part is of the form $\exp [-S+\text{(extra terms)}]$ ($S$ be the ``Euclidean" action $(29)$) because of the term $\ll$angular momentum$\gg$.

According to the usual method, in order to introduce the interaction, let us define the generating functional $Z_0[J,J^{\dagger}]$ for free theory.

\begin{eqnarray}
	& & Z_0[J,J^{\dagger}]\equiv \int \mathfrak{D}\phi \mathfrak{D}
	\phi^{\dagger} \exp \big[ -\int d^2t \phi^{\dagger} \Big\{ 
	\frac{1}{\theta}(\hat{N}+1)+\frac{1}{\theta}\hat{L}+m^2
	+\hat{M}\Big\} \phi \nonumber \\
	& & \qquad \qquad \qquad \qquad \qquad \qquad \qquad \qquad \qquad 
	+\int d^2t (J\phi +J^{\dagger}\phi^{\dagger})\big] \\
	& & \qquad \qquad \hat{M} \equiv -\frac{1}{2\theta^2}t_2^2+(1+i)
	\frac{i}{\theta}t_2\partial_1
\end{eqnarray}

$Z_0[J,J^{\dagger}]$ is supposed to be normalized such that $Z_0[J=0,J^{\dagger}=0]=1$. The term $\displaystyle \big[-\int d^2t \phi^{\dagger}\hat{M}\phi \big]$ is the (extra terms) mentioned above. Before turning to a explicit evaluation of $Z_0[J,J^{\dagger}]$, the operator $\hat{M}$ must be clarified. As we shall see in the following calculation, the operator $\hat{M}$ can be expressed with the ``raising and lowering" operators $\hat{N}_{\pm},\hat{N}_{\pm}^{\dagger}$. From $(33)(34)$, we obtain

\begin{eqnarray*}
	& & \qquad t_2=\frac{1}{2}\theta \big\{ (\hat{N}_{+}+\hat{N}_{-})
	-(\hat{N}_{+}^{\dagger}+\hat{N}_{-}^{\dagger})\big\} \\
	& & \qquad \partial_1 =-\frac{i}{4}\big\{ (\hat{N}_{+}+\hat{N}_{-})
	-(\hat{N}_{+}^{\dagger}+\hat{N}_{-}^{\dagger})\big\} \text{.}
\end{eqnarray*}

Thus,

\begin{eqnarray}
	\hat{M} \! \! \! \! \! \! \! \! \! & & =-\frac{1}{2\theta^2}t_2^2
	+(1+i)\frac{i}{\theta}t_2 \partial_1 \nonumber \\
	& & =-\frac{1}{8}\big\{ (\hat{N}_{+}+\hat{N}_{-})
	-(\hat{N}_{+}^{\dagger}
	+\hat{N}_{-}^{\dagger})\big\}^2-\frac{1}{8}(1+i)\big\{ (\hat{N}_{+}
	+\hat{N}_{-})^2-(\hat{N}_{+}^{\dagger}+\hat{N}_{-}^{\dagger})^2 \big\} 
	\nonumber \\
	& & =-\frac{1}{8}[2\big\{ (\hat{N}_{+}^{\dagger}
	+\hat{N}_{-}^{\dagger})^2-(\hat{N}_{+}+\hat{N}_{-})
	(\hat{N}_{+}^{\dagger}+\hat{N}_{-}^{\dagger})\big\} -i\big\{ 
	(\hat{N}_{+}+\hat{N}_{-})^2-(\hat{N}_{+}^{\dagger}
	+\hat{N}_{-}^{\dagger})^2 \big\} ]\text{,} \nonumber \\
	\langle N,l|\! \! \! \! \! \! \! \! \! \! & & 
	\hat{M} |N^{\prime},l^{\prime}\rangle \nonumber \\
	& & =-\langle N,l|\big\{ \frac{1}{4}(\hat{N}_{+}^{\dagger}
	\hat{N}_{-}^{\dagger}+\hat{N}_{-}^{\dagger}\hat{N}_{+}^{\dagger})
	+\frac{i}{8}(\hat{N}_{+}^{\dagger}\hat{N}_{-}^{\dagger}
	+\hat{N}_{-}^{\dagger}\hat{N}_{+}^{\dagger})-\frac{i}{8}(\hat{N}_{+}
	\hat{N}_{-}+\hat{N}_{-}\hat{N}_{+})\big\} |N^{\prime},l^{\prime}\rangle
	 \nonumber \\
	& & \qquad \qquad \qquad +\text{(remainders)} \nonumber \\
	& & =-\frac{1}{2}\big\{ \frac{1}{\theta}(N+1)+\frac{1}{\theta}l\big\} 
	\delta_{NN^{\prime}}\delta_{ll^{\prime}}-\frac{i}{2}\frac{1}{\theta}l
	\delta_{NN^{\prime}}\delta_{ll^{\prime}}+\text{(remainders)} \text{.}
\end{eqnarray}

Here, (remainders) are the terms which are not proportional to $\delta_{NN^{\prime}}\delta_{ll^{\prime}}$ ,such as \[ \langle N,l|\hat{N}_{+}^{\dagger}\hat{N}_{+}^{\dagger}|N^{\prime},l^{\prime}\rangle \propto \delta_{N,N^{\prime}+2}\delta_{l,l^{\prime}+2}\text{.}\]

Now that we have got the necessary information, let us evaluate the generating functional $Z_0[J,J^{\dagger}]$. From this evaluation, we will obtain the propagator, and be able to introduce the multi-point ``local" interaction. Although the practical calculation is slightly tedious, we shall show it in detail.

\begin{eqnarray}
	Z_0[J,J^{\dagger}] \! \! \! \! \! \! \! \! \! & & =\int \mathfrak{D}
	\phi \mathfrak{D}\phi^{\dagger} \exp \big[ -\! \! \int d^2t 
	\phi^{\dagger} \Big\{ \frac{1}{\theta}(\hat{N}+1)+\frac{1}{\theta}
	\hat{L}+m^2+\hat{M}\Big\} \phi +\! \! \int d^2t (J\phi +J^{\dagger}
	\phi^{\dagger})\big] \nonumber \\
	& & \qquad \qquad \text{We expand as } \left\{ \begin{array}{l}
	\phi (t_1,t_2)=\displaystyle \sum_{N,l} \alpha_{N,l}\langle t_1,t_2|N,l
	\rangle \\
	\phi^{\dagger}(t_1,t_2)=\displaystyle \sum_{N,l} \alpha^{\dagger}_{N,l}
	\langle N,l|t_1,t_2\rangle \\
	J(t_1,t_2)=\displaystyle \sum_{N,l} J_{N,l}\langle t_1,t_2|N,l\rangle 
	\\
	J^{\dagger}(t_1,t_2)=\displaystyle \sum_{N,l} J^{\dagger}_{N,l}\langle 
	t_1,t_2|N,l\rangle 
	\end{array} \right. \nonumber \text{,} \\
	& & =\prod_{N=0}^{\infty}\prod_{-N\leq l\leq N}\int d\alpha_{N,l}
	d\alpha^{\dagger}_{N,l} \nonumber \\
	& & \qquad \qquad \times \exp \big[ -\sum_{N=0}^{\infty}
	\sum_{-N\leq l\leq N}\Big\{ \frac{1}{2\theta}(N+1)+\frac{1}{2\theta}l
	+m^2-\frac{i}{2\theta}l\Big\} \alpha^{\dagger}_{N,l}\alpha_{N,l} 
	\nonumber \\
	& & \qquad \qquad +\text{(remainders)}+\sum_{N=0}^{\infty}
	\sum_{-N\leq l\leq N}\big( J_{N,l}\alpha_{N,l}+J^{\dagger}_{N,l}
	\alpha^{\dagger}_{N,l}\big) \big] \nonumber \\
	& & \qquad \left( \begin{array}{l}
	\text{The terms (remainders) come from the term $\phi^{\dagger} 
	\hat{M} \phi$.} \\
	\text{Let us carry out the (remainders)' integration symbolically.} \\
	\int d\alpha_Nd\alpha^{\dagger}_N \exp [-(\alpha^{\dagger}_N\alpha_N+
	\alpha^{\dagger}_N\alpha_{N+2})] ,\; \; \alpha_N\equiv x_N+iy_N \\
	\sim \int dx_Ndy_N \exp [-(x_N^2+y_N^2) \\
	\qquad \qquad \qquad \qquad \qquad +(x_{N+2}+iy_{N+2})x_N-i(x_{N+2}
	+iy_{N+2})y_N] \\
	\sim \exp [\frac{(x_{N+2}+iy_{N+2})^2}{4}
	-\frac{(x_{N+2}+iy_{N+2})^2}{4}] \\
	\sim 1
	\end{array} \right) \nonumber \\
	& & =\exp \Big[ \sum_{N=0}^{\infty}\sum_{-N\leq l\leq N}
	J^{\dagger}_{N,l}\Big\{ \frac{1}{2\theta}(N+1)+\frac{1}{2\theta}l+m^2
	-\frac{i}{2\theta}l\Big\}^{-1}J_{N,l} \Big]
\end{eqnarray}

From this, we find that the propagator in ``momentum" space $(N,l)$ is 

\begin{eqnarray}
	\Big\{ \frac{1}{2\theta}(N+1)+\frac{1}{2\theta}l+m^2
	-\frac{i}{2\theta}l\Big\}^{-1} \text{.}
\end{eqnarray}

The generating functional $Z[J,J^{\dagger}]$ for interacting theory can be written in the same way.

\begin{eqnarray}
	& & Z[J,J^{\dagger}]=\int \mathfrak{D}\phi \mathfrak{D}\phi^{\dagger} 
	\exp \Big[ -S-\! \! \int d^2t \phi^{\dagger}\hat{M}\phi -\! \! \int 
	d^2t \mathfrak{L}_{\text{int}}[\phi ,\phi^{\dagger}] +\! \! \int d^2t 
	(J\phi +J^{\dagger}\phi^{\dagger})\Big] \nonumber \\
	& & \qquad \quad \Big( \text{for example, } \quad 
	\mathfrak{L}_{\text{int}}=\frac{\lambda}{4}(\phi^{\dagger})^2(\phi )^2 
	\Big)
\end{eqnarray}

The Feynman rule is much the same as the usual interacting scalar field theory, except giving the lines the propagator $(70)$ and taking summation $(N,l)$ for the inner lines.

As is well known, the usual scalar field theory has the logarithmic UV divergences, but I do not know about $(71)$ enough to be able to say that it's amplitudes diverge or converge, yet. As the spectrum $(N,l)$ is discrete, however, the degree of divergence is expected to be lowered, or the existence of the imaginary part in the propagator $(70)$ may make the loop amplitudes convergent.

As the last comment of this section, let us refer to the ``Minkowskian" functional integral. Starting from the action $(52)$ and using the ``real-time" evolution operator, we will arrive at the functional integral of the integrand $\exp [-S_M]$ after a similar calculation. This is, in a sense, the preferable result. But, instead of the strange functional integral in the ``Euclidean" case, it is difficult to find the complete system which expands $\phi$ and $\phi^{\dagger}$ in the action $S_M$, in other words, difficult to first-quantize the system $S_M$. In future, I would like to examine the action $S_M$ closely.

\vskip 1.5cm

\noindent
{\bf 6. Discussions and acknowledgements}

\vskip 0.5cm

In this paper, we constructed the model of the scalar field theory on the noncommutative space from the point of view that the first-quantization is equal to making the space-time noncommutative and that the first-quantized quantum mechanics is equal to the classical field theory. We carried out the first, second quantizations. The advantage of our formulation is that the method is very similar to that of the usual scalar field theories contrary to the approach og using the $*$-product. The latter has the logarithmic UV divergences and also the IR ones. Our main results are summarized in the following two. One is that the spectrum is discrete given by $(N,l)$. The other is that the propagator $(70)$ has the imaginary part which looks like the regulator. Thus we may expect the loop amplitudes convergent. But investigation of the symmetries and conserved quantities are left to the future works. As our action $S$ in $(29)$ manifestly depends on the space coordinates $t_1,t_2$, it doesn't keep the symmetry of translation. Therefore, the usual energy-momentum is not conserved. But the action $S$ maintains the rotational invariance in the space $(t_1,t_2)$, and the relation between this rotational symmetry and the real Lorentz invariance should be clarified.

Although, for simplicity, we have used the ``Euclidean" noncommutative space, and got the ``Euclidean" functional integral with the ``Minkowskian" system which evolves in the ``imaginary-time", we can discuss in the ``Minkowskian" metric instead of ``Euclidean" equation $(12)$. In this case, the first-quantization cannot be done by the harmonic oscillator system, and one cannot say something definite about the spectrum. The functional integral by the evolution in the ``real" time, however, becomes the familiar form such as $\int \mathfrak{D}\phi \exp [-S_M]$. So, if we can find the complete system of the first-quantized theory, it may be easy to treat the functional integral.

As a interesting extension, we can also impose the noncommutativity on the fields $\phi$ and $\phi^{\dagger}$ such as 

\begin{eqnarray*}
	& & [\phi (t_1,t_2),\phi (t_1,t_2^{\prime})]=i\theta(t_2-t_2^{\prime}) 
	\\
	& & [\phi^{\dagger} (t_1,t_2),\phi^{\dagger} (t_1,t_2^{\prime})]
	=-i\theta(t_2-t_2^{\prime}) \qquad \text{etc.} \\
	& & \theta (x)=
	\left\{ \begin{array}{ll}
	+1 & ,x>0 \\
	0 & ,x=0 \\
	-1 & ,x<0
	\end{array} \right.
\end{eqnarray*}

\noindent and, I think we must incorporate this commutation relation into the second-quantization, because the amplitude has only 4-excitation states (see $(45)\sim (48)$).

This noncommutativity will lead us to the strange ``creation-annihilation" algebra and functional integral which also fairly differ from $(45)\sim (48)$ and $(65)$, and produce another effects on the convergence of amplitudes.

As the other future works, the introduction of the gauge interactions may be interesting. As is well known (see, for example, Ref.~\cite{landi}), the gauge potential and the connection can be introduce into the context of the noncommutative geometry. From the point of view that the first-quantization is by itself the construction of the field theories, our new method may be needed for the noncommutative version of Yang-Mills theory and gravity model. At least, the degree of divergence seems to be lowered in our model, so the application of our method to the other field theories may be useful.

\vskip 0.5cm

Finally, I would like to thank M.Ninomiya for his continuous encouragement and carefully reading the manuscript. I also would like to acknowledge my colleagues for useful comments and discussions.

\end{document}